\documentclass[aps,prl,amsmath,twocolumn,showpacs,superscriptaddress]{revtex4}
\usepackage{graphicx}

\begin{document}

\title{Optical spin orientation of a single manganese atom in a quantum dot}

\author{C. Le Gall}
\author{L. Besombes}
\email{lucien.besombes@grenoble.cnrs.fr}
\author{H. Boukari}
\author{R. Kolodka}
\author{H. Mariette}
\author{J. Cibert}
\affiliation{Institut N\'eel, CNRS \& Universit\'e Joseph Fourier,
25 avenue des Martyrs, 38042 Grenoble, France}

\date{\today}

\begin{abstract}

A hight degree of spin polarization is achieved for a Mn atom
localized in a semiconductor quantum dot using quasi-resonant
optical excitation at zero magnetic field. Optically created spin
polarized carriers generate an energy splitting of the Mn spin and
enable magnetic moment orientation controlled by the photon helicity
and energy. The dynamics and the magnetic field dependence of the
optical pumping mechanism shows that the spin lifetime of an
isolated Mn atom at zero magnetic field is controlled by a magnetic
anisotropy induced by the built-in strain in the quantum dots.

\end{abstract}

\pacs{78.67.Hc, 78.55.Et, 75.75.+a}

\maketitle

Controlling the interaction between spin polarized carriers and
magnetic atoms is of fundamental interest to understand the
mechanism of spin transfer. Spin transfer could lead to the
development of devices in which the spin state of magnetic atoms is
controlled by the injection of a spin polarized current and not by
an external magnetic field, as in conventional magnetic memories.
Information storage on a single magnetic atom would be the ultimate
limit in the miniaturisation of these magnetic memories. The
performance of such a device will rely on the lifetime of the
isolated spin. Dilute magnetic semiconductors (DMS) systems
combining high quality semiconductor heterostructures and the
magnetic properties of Mn dopant are good candidates for these
ultimate devices \cite{Fernandez2007}. In a DMS, spin polarized
carriers introduced by optical excitation couple strongly with
localized Mn spins allowing, \emph{e.g.}, photo-induced
magnetization to be achieved \cite{Krenn1989,kudinov2003}. It has
been known for years that, in the absence of carriers and under
magnetic field, highly dilute ensembles of Mn present a spin
relaxation time in the millisecond range \cite{dietl95}.  However,
the dynamics of such ensembles can be much faster at zero field
\cite{Goryca2008}. In addition, very few is known about the zero
field dynamics of a single Mn spin, and the optical spin orientation
of a single magnetic atom in a solid state environment is still a
challenge \cite{Myers2008}.

We demonstrate a new way to optically address the spin of a single
Mn atom localized in an individual quantum dot (QD): the injection
of spin polarized carriers under selective optical excitation is
used to prepare a non-equilibrium Mn spin distribution without any
applied magnetic field. Photoluminescence (PL) transients recorded
when switching the circular polarization of the excitation reflect
the dynamics of this optical orientation mechanism. The magnetic
field dependence of the optical pumping efficiency reveals the
influence of the Mn fine structure on the spin dynamics: the strain
induced magnetic anisotropy of the Mn spin slows down the relaxation
at zero magnetic field. We show that the Mn spin distribution
prepared by optical pumping is fully conserved for a few
microseconds.

Growth and optical addressing of QDs containing a single magnetic
atom were achieved recently
\cite{Besombes04,Maingault2006,Kudelski2007}. The static properties
of these systems are now well understood: when a Mn dopant atom is
included in a II-VI QD, the spin of the optically created
electron-hole pair interacts with the five {\it d} electrons of Mn
(total spin $S$=5/2). This leads to a splitting of the once simple
PL spectrum of an individual QD into six ($2S+1$) components, as
shown in the bottom of Fig.~1(a). This splitting results from the
spin structure of the confined holes which are quantized along the
QDs' growth axis with their spin component taking only the values
$J_z=\pm3/2$ \cite{Leger2005}. The hole-Mn exchange interaction
reduces to an Ising term $J_z S_z$ and shifts the PL energy of the
QD according to the relative projection of the Mn and hole spins
\cite{Leger2007}. The intensity of each line reflects the
probability for the Mn to be in one of its six spin components and
it is a probe of the Mn spin state at the moment the exciton
recombines \cite{Besombes2008}. As the spin state of the Mn
fluctuates during the optical measurements, the six lines are
observed in the time averaged PL spectrum.

To optically pump the Mn spin, the PL of a single QD was
quasi-resonantly excited with a tunable continuous wave (CW) dye
laser. In order to record the spin transients, the linear
polarization of the excitation laser was modulated between two
orthogonal states by switching an electro-optic modulator with a
rise time of 5~ns, and converted to circular polarization with a
quarter-wave plate. Trains of resonant light with variable duration
were generated from the CW laser using acousto-optical modulators
with a switching time of 10~ns. The circularly polarized collected
light was dispersed by a 1~m double monochromator before being
detected by a fast avalanche photodiode in conjunction with a time
correlated photon counting unit with an overall time resolution
$\sim$~40~ps.

\begin{figure}[bt]
\includegraphics[width=2.8in]{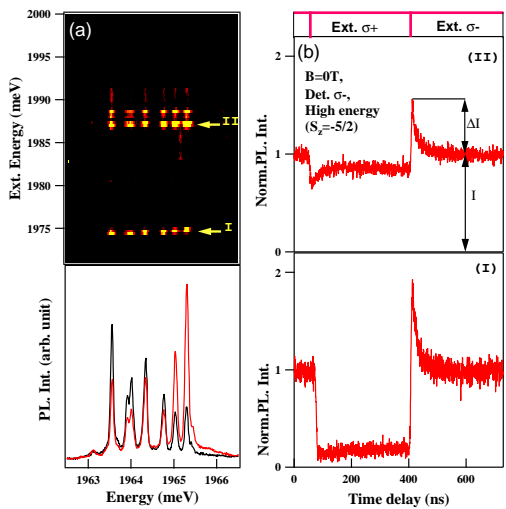}
\caption{(Color online) (a) PL and PL excitation spectra of a single
Mn-doped QD at $B=0$~T and $T=5$~K. The PL is detected in circular
polarization under an alternate $\sigma^- / \sigma^+$ excitation at
two different wavelengths into the same set of excited states: at
1987.0~meV (black) and 1987.4~meV (red). (b) PL transient under
polarization switching at $B=0$~T. PL is detected on the high energy
line of X-Mn in $\sigma^-$ polarization (Mn spin $S_z=-5/2$).
Transient (I) (resp. (II)) was observed under resonant excitation at
1975~meV (resp. 1987~meV)} \label{fig1}
\end{figure}

Fig.~1 and 2 summarize the main features of the time-resolved
optical orientation experiment. The PL of the exciton-Mn (X-Mn)
system was excited in one of the sharp lines observed at slightly
higher energy (top of Fig.~1(a)), \emph{i.e.}, in an excited states
of the X-Mn complex \cite{Glazov07}; the PL intensity was measured
in circular polarization (\emph{e.g.}, $\sigma^-$, corresponding to
the recombination of the $-1$ exciton). Then, the relative intensity
of the six lines dramatically depends on the excited state selected
for the excitation (bottom of Fig.~1(a)): as each line corresponds
to one value of the Mn spin projection $S_z$, that shows that the
whole process creates a non-equilibrium distribution of the Mn spin
states in the ground state of X-Mn. Under these conditions,
switching the circular polarization of the excitation produces a
change of the PL intensity (Fig.~1(b)) with two transients: first an
abrupt one with the same sign for all six lines, reflecting the
population change of the spin polarized excitons; then a slower
transient with opposite signs on the two extreme PL lines
(\emph{i.e.}, when monitoring the Mn spin states $S_z=5/2$ or
$S_z=-5/2$, Fig.~2(b)), and
 a characteristic time which is inversely proportional to
the pump intensity (Fig.~2(a)).

This is the signature of an optical pumping process which realizes a
spin orientation of the Mn impurity. We first discuss the details of
this process, then use it to study the spin dynamics of the single
Mn in the QD.

The relevant sublevels of X-Mn and Mn are schematized in Fig.~2(c).
For the sake of simplicity, we omit the dark exciton states which
should be included for a quantitative analysis. When exciting one of
the low energy excited states of the QD, two mechanisms are expected
to contribute to the observed spin orientation \cite{Glazov07}: the
selective excitation of the QD can show a strong dependence on the
Mn spin state, and the relaxation of the Mn spin within the X-Mn
system is driven by the interaction with the spin polarized carriers
which have been injected.

Under spin selective excitation, spin relaxation of X-Mn tends to
empty the spin state of the Mn which gives rise to an absorption
maximum \cite{govorov2005}.

Within the X-Mn complex, the spin-flip time $\tau_{X-Mn}$ is
influenced by carrier-phonon, carrier-nuclei and exchange
interactions affecting the exciton \cite{govorov2005,feng2007}. In
particular, the hole spin relaxation is controlled by the
interaction with phonons \cite{woods2004} and can be in the ns range
resulting in a partial thermalization of the X-Mn complex
\cite{Besombes2008}. The off-diagonal terms of the $sp-d$ exchange
interaction \cite{Leger2007,Besombes2008} allow simultaneous
spin-flips of carrier and Mn spins, and a spin transfer from the
injected carriers to the localized $d$ electrons of Mn is made
possible. If these X-Mn spin-flips are faster than the spin
relaxation of the Mn alone, a dynamic spin orientation of the Mn can
be performed. Under injection of spin polarized carriers, this
relaxation process tends to anti-align the Mn spin with the exchange
field of the exciton to reach a thermal equilibrium on the X-Mn
levels \cite{govorov2005}. Hence, optical pumping with $\sigma^-$
photons for instance, tends to decrease the population of the spin
state S$_z$=-5/2 and decrease that of S$_z$=+5/2, as observed in
Fig.2(b).

Both mechanisms, absorption selectivity and spin injection, depend
on the structure of the excited states, resulting in a pumping
signal which depends on the excitation energy (Fig.~1). An efficient
pumping of the Mn spin can be performed within a few tens of
nanoseconds, showing that at $B=0$ the spin relaxation time of the
Mn alone is long enough compared to the X-Mn dynamics.

\begin{figure}[bt]
\includegraphics[width=2.8in]{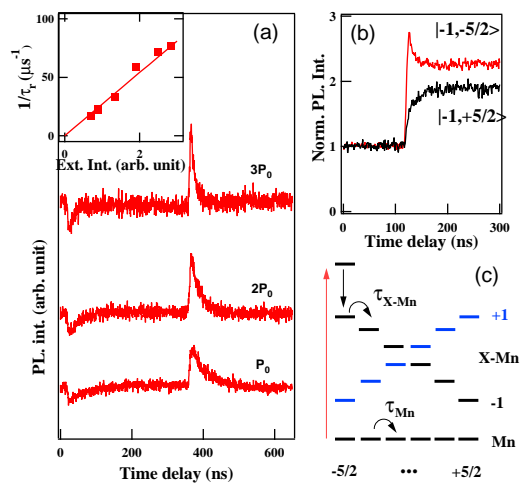}
\caption{(Color online) PL transients at different values of the
excitation power. Inset: power dependence of the inverse response
time $\tau_r$, taken at the $1/e$ point of the spin-related
transient. (b) PL transients recorded in $\sigma^-$ polarization on
the high ($S_z$=-5/2) and low ($S_z$=+5/2) energy line of the X-Mn
complex. (c) Simplified level diagram of a Mn-doped QD, as a
function of Mn spin (X-Mn: bright exciton-Mn).} \label{fig2}
\end{figure}

\begin{figure}[bt]
\includegraphics[width=2.8in]{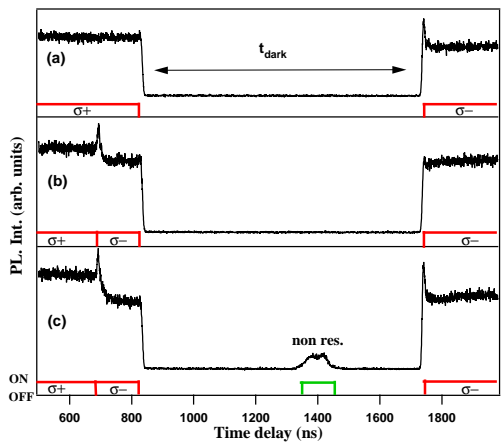}
\caption{ (Color online) PL transients recorded under the optical
polarization sequence displayed at the bottom of each plot. The spin
distribution prepared by optical pumping is conserved during the
dark time $t_{\textrm{dark}}$: in (a) and (b), a transient is
observed only for a different helicity of the pump and the probe (in
(a), the polarization is switched during the dark time). The
injection of high energy carriers with a non-resonant excitation
(514~nm) forces the relaxation of the Mn spin: the optical pumping
appears after $t_{\textrm{dark}}$ (c).} \label{fig3}
\end{figure}

Having established a method to prepare Mn spins, we performed
pump-probe experiments to observe how the Mn polarization is
conserved (Fig.~3). We prepare a non-equilibrium distribution of the
Mn spin with a $\sigma^\pm$ pulse. The pump laser is then switched
off, and switched on again after a dark time $t_{\textrm{dark}}$. We
observe no transient if the laser is switched on with the same
polarization, and a transient of constant intensity when the
polarization has been changed, whatever the value of
$t_{\textrm{dark}}$ up to the $\mu$s range. This demonstrates a
complete conservation of the prepared Mn spin distribution over
microseconds.

In addition, as expected from previous measurements of a single Mn
spin dynamics under CW optical excitation \cite{Besombes2008}, the
injection of high energy carriers in the vicinity of the QD
significantly increases the Mn spin relaxation rate. This is
illustrated in Fig.~3(c): the PL transient is recovered when free
carriers have been injected during the dark time with a second
non-resonant laser, thus erasing the single spin memory.

More information can be obtained from the magnetic field dependence
of the optical pumping signal. For an isotropic Mn spin, the
decoherence of the precessing spin gives rise to the standard Hanle
depolarization curve with a Lorentzian shape and a width
proportional to $1/T_2$ \cite{Myers2008}. In the present case, a
magnetic field in the Faraday configuration ($B_z$) does not change
significantly the PL transients (Fig.~4(b)): a weak increase of the
spin orientation efficiency is observed as soon as a field of a few
mT is applied; above that, the pumping efficiency remains constant.
By contrast, an in-plane field ($B_x$) induces coherent precession
of the Mn spin away from the optical axis (=~QDs' growth axis) , so
that the average spin polarization, and therefore the amplitude of
the optical pumping signal, decay (Fig.~4(a)).

\begin{figure}[bt]
\includegraphics[width=2.8in]{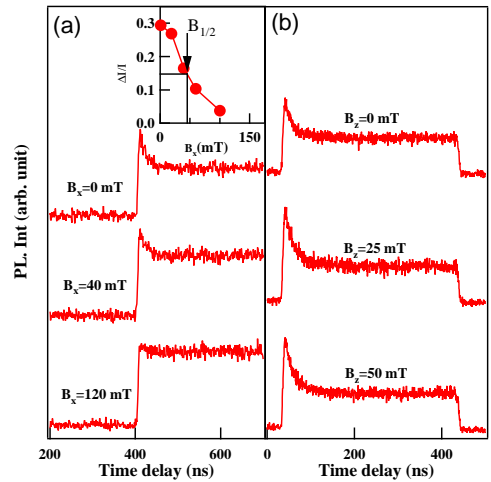}
\caption{ (Color online) Mn spin transient as a function of a
magnetic field applied in-plane (a) and out-of-plane (b). The inset
in (a) shows the dependence of the transient amplitude $\Delta I/I$
(defined in Fig.~1) on the transverse field . $B_{1/2}$ corresponds
to the half width at half maximum. } \label{fig4}
\end{figure}

It is known from electron paramagnetic resonance spectroscopy that
the ground state of Mn$^{2+}$ presents a fine structure
\cite{Furdyna95}. In a cubic crystal, it results from a strong
hyperfine coupling with the Mn nuclear spin,
$A$~\textbf{I}.\textbf{S} (with $I=5/2$ and $A\approx0.7$$\mu$eV),
and the crystal field. In addition, in epitaxial structures,
built-in strains due to the lattice mismatch induce a magnetic
anisotropy with an easy axis along the QD axis; it scales as
$D_0S_z^2$, with $D_0$ proportional to the tetragonal strain. The
resulting fine structure under a magnetic field applied in-plane or
out-of-plane is shown in Fig.~5(a). At zero field, the Mn electronic
spin is quantized along the growth axis and the different spin
doublets are split by an energy proportional to ${D_0}$.

\begin{figure}[bt]
\includegraphics[width=2.8in]{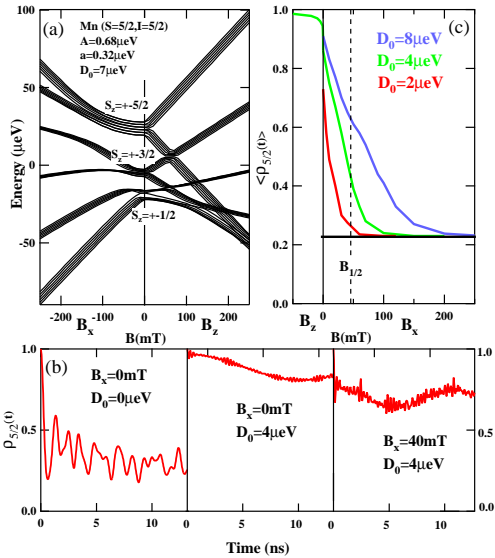}
\caption{ (Color online) (a) Magnetic field dependence of the fine
structure of the Mn spin with out-of-plane (right) and in-plane
(left) field, calculated with $A=0.68~\mu$eV, $D_0=7~\mu$eV and a
cristal field parameter $a=0.32~\mu$eV. (b) Time evolution of the
density matrix element $\rho_{+5/2}(t)$ calculated for different
values of $D_0$ and of the transverse magnetic field $B_x$
($\rho_{+5/2}(0)=1$). (c) Time average value of
$\langle\rho_{+5/2}(t)\rangle$ versus $B_x$ for different values of
$D_0$ ($T_2=\infty$). The $B_{1/2}$ found experimentally is
indicated with a dotted line.}\label{fig5}
\end{figure}

This fine structure controls the Mn spin dynamics at zero or weak
magnetic field. At zero field, in the absence of anisotropy, the
precession of the electronic spin of Mn in its own hyperfine field
should erase any information stored on the electronic spin
\cite{Goryca2008}. In the presence of anisotropy, the precession of
the Mn spin in the nuclear field is blocked even at $B=0$. This is
shown in Fig.~5(b), where we plot the evolution of the density
matrix element $\rho_{5/2}(t)$, calculated for the
electronic+nuclear spin system of a single Mn ion with no relaxation
(\emph{i.e}., no coupling to the environment), with the initial
condition $\rho_{5/2}(0)=1$. With $D_0=0$, the coherent evolution of
the electronic spin in the nuclear field erases the orientation of
the electronic spin in a few 100~ps. With a $D_0$ of a few $\mu$eV,
this free precession is blocked and the Mn spin state can be
conserved for a long time, in agreement with Fig.~3. In particular,
it is conserved during the time interval between the injection of
two consecutive excitons, allowing the optical pumping mechanism to
take place.

The magnetic anisotropy also blocks the Mn spin precession in a weak
transverse magnetic field. Then the field dependence of the optical
pumping efficiency is controlled by the anisotropy $D_0$ and the
coherence time $T_2$. A dephasing time $T_2^*\simeq$1~ns has been
measured in an ensemble of CdMn(2\%)Se QDs \cite{Scheibner06}: a
longer decoherence time $T_2$ is expected for a single isolated Mn
spin. For $T_2 \geq$1~ns, the influence of the decoherence on the
width of the depolarization curve is smaller than 4~mT and its
contribution to the experimental curve (Fig.~3(a)) can be neglected.
Hence, in order to estimate $D_0$, we take again $T_2=\infty$ to
calculate the density matrix element $\rho_{5/2}(t)$ for
$\rho_{5/2}(0)=1$, assuming various values of $D_0$.

Fig.~5(c) displays the magnetic field dependence of the
time-averaged value of $\rho_{5/2}(t)$. This quantity describes the
probability for the state $S_z=+5/2$ to be conserved after the
recombination of an exciton, as the electronic Mn spin evolves in
the hyperfine field, the crystal field and the applied magnetic
field. A decrease in this spin conservation progressively destroys
the cumulative process controlling the optical pumping mechanism.
For a free precessing spin, $\langle\rho_{5/2}(t)\rangle\approx0.24$
as soon as a transverse field is applied and the depolarization is
controlled by $T_2$ \cite{Myers2008}. In the presence of anisotropy,
the Mn spin does not precess at weak field and the Mn spin state is
partially conserved (Fig.~5(b)). When the transverse magnetic field
is strong enough to overcome the magnetic anisotropy
($g_{\textrm{Mn}} \mu_{\textrm{B}}B_x\gg D_0$), the time average of
$\rho_{5/2}(t)$ reaches the expected value for a coherently
precessing spin. This progressive decrease of
$\langle\rho_{5/2}(t)\rangle$ gives rise to a field-induced
depolarization curve which depends on the value of $D_0$. A
half-height field $B_{1/2}\simeq$45~mT, as observed experimentally
in Fig.~4(b), is obtained for $D_0\approx6$~$\mu$eV. As a value
$D_0\approx12$~$\mu$eV is expected for CdTe coherently grown on ZnTe
\cite{Furdyna95}, this is consistent with the presence of a CdZnTe
alloy or a partial relaxation of the mismatch strain at the Mn
location.

A  magnetic anisotropy large enough to block most of the Mn spin
relaxation also explains the very weak influence of an out-of-plane
field. Nevertheless, the Zeeman splitting cancels the residual
non-diagonal coupling induced by crystal field and slightly improves
the Mn spin conservation (left part in Fig.~5(c)), thus accounting
for the small increase of the  optical pumping efficiency
experimentally observed in Fig.~4(b).

To conclude, our results demonstrate the zero-field optical spin
orientation of a single magnetic atom in a semiconductor host.
Selective optical excitation of an individual Mn-doped QD with
circularly polarized photons can be used to prepare a
non-equilibrium distribution of the Mn spin without any applied
magnetic field. Fully resonant excitation of the Mn-doped QD ground
state should permit a higher fidelity preparation of the Mn spin.
With this new preparation and readout scheme it should be possible
to initialize a Mn spin, manipulate it on a microseconds time-scale
with resonant microwave excitation and reliably read the final
state.

This work was supported by French ANR contracts MOMES and CoSin.

\end{document}